# Molecular Transport Junctions: Propensity Rules for Inelastic Electron Tunneling Spectra

Alessandro Troisi[a)] and Mark A. Ratner[b)]

[a)] *Department of Chemistry, University of Warwick, CV4 7AL Coventry, UK*

[b)] *Department of Chemistry, Center for Nanofabrication and Molecular Self-Assembly, Northwestern University, Evanston, Illinois 60208*

*Abstract*

We develop a series of propensity rules for interpreting Inelastic Electron Tunneling (IET) spectra of single-molecule transport junctions. IETS has no selection rules, such as those seen in optical, infrared and Raman spectra, because IETS features arise not from the field-dipole interaction characterizing these other spectroscopies, but from vibronic modification of the electronic levels. Expansion of the Landauer-Imry formula in Taylor series in molecular normal coordinates gives a convenient, accurate perturbation-type formula for calculating both frequency and intensity of the IETS spectrum. Expansion in a Dyson-like form permits derivation of propensity rules, both symmetry-based and pathway-deduced, allowing correlation of structure and coupling geometry with the IETS spectrum. These propensity rules work very well for the calculated spectrum of four typical molecular bridges.

Molecular transport junctions (MJ) contain few or single molecules. Although their transport properties are being addressed extensively, the community still has no effective structural probe. For the standard metal/molecule/metal (almost always the metal is gold) junction, this is serious, since slightly differing contact geometries yield different values of the conductance. Since we have little idea of the molecular geometry in the junction, especially under current, possible molecular rearrangement features cannot be directly examined.

Inelastic electron tunneling spectroscopy (IETS) measures peaks in the second derivative of current $d^2I/d\Phi^2$ (where $\Phi$ and $I$ are respectively potential and current) and has been of interest in microscopic structures for half a century.[1-3] Their observation in MJ is newer,[4-6] and may present the first hope for systematically deducing any structural information concerning the molecule in the junction. We present here a derivation of propensity rules for IETS in MJ, based on use of perturbation theory. These rules describe which molecular vibrations should dominate, for given binding geometry and molecular configuration, and should complement other purely computational approaches.[7-15]

IETS occurs at low voltages, in the tunneling-type Landauer-Imry regime.[16,17] Timescale considerations suggest that vibronic coupling effects should be weak, because the time of contact between tunneling electrons and molecule is on the femtosecond scale or faster. Measurements of IETS bear this out—no sharp structures are seen directly in the I-V curve in this region, because the inelastic channel is very weak compared to the elastic one. The IETS spectrum shows peaks, occurring very near the frequencies of particular normal modes of the parent molecule. Our purpose is to derive propensity rules describing the IETS spectrum, facilitating its assignment and, therefore, the junction characterization.

***Background***. Because IETS is so weak, it is reasonable to use perturbation theory. In the tunneling regime, the Landauer-Imry equation for the conductance can be written

$$g^{el}(E) = g_c\,\mathrm{Tr}(\mathbf{\Gamma}^L(E)\mathbf{G}(E)\mathbf{\Gamma}^R(E)\mathbf{G}(E)^+) \qquad (1)$$

Here $E$, $\mathbf{\Gamma}^{L(R)}$, $\mathbf{G}$, $g_c$, are respectively the energy variable, the spectral density coupling the molecular ends with the left (or right) electrode, the retarded Green's function describing charge propagation within the molecule, and the conductance quantum (1/12.8kΩ).[16] In the IETS regime, which occurs at low $\Phi$, we can replace the energy variable by its value at the Fermi energy, $E_F$. A peak due to the normal mode $Q_\alpha$ has integrated intensity (area below the peak) $W_\alpha$:[10]

$$W_\alpha = g_c \mathrm{Tr}(\mathbf{\Gamma}^L(E_F)\mathbf{G}^\alpha(E_F)\mathbf{\Gamma}^R(E_F)\mathbf{G}^\alpha(E_F)^+) \quad (2)$$

Here Taylor expansion has been used (like the Herzberg-Teller expansion of molecular vibronics), and the notation is defined as $G^\alpha_{ij} = \sqrt{2}/2\left(\partial G_{ij}(E,\{Q_\alpha\})/\partial Q_\alpha\right)_{\{Q_\alpha\}=0}$ — that is, it is the derivative of the Green's function for propagation with respect to the normal (dimensionless) molecular coordinate $Q_\alpha$.

The spectral density $\mathbf{\Gamma}$ arises from the lead/molecule coupling, with matrix elements $\Gamma^L_{ij}(E) = -2\pi\sum_\lambda V_{\lambda i}V_{\lambda j}{}^*\delta(E_\lambda - E)$; indexes *i,j* run over the local orbitals of the molecule, and $\lambda$ over the the eigenstates (of energy $E_\lambda$) of the metallic electrode. $V_{\lambda i}$ is the Hamiltonian matrix element between atomic orbital *i* on the molecule and eigenstate $\lambda$ in the metal.

We assume that the Γ's are dominated by a few local orbitals on the interface atoms of the molecule (for instance, the sulfur atomic orbitals in gold/thiol junctions, or nitrogen orbitals in CN/Pd junctions) distinct for the two electrodes. Then $\Gamma^L$ and $\Gamma^R$ can be diagonalized by a unique transformation, giving for the overall conduction

$$g(E) = g_c \sum_{\substack{m\in\mathcal{L}\\ n\in\mathcal{R}}} \Gamma^L_{mm}(E)\Gamma^R_{nn}(E)\left|G_{mn}(E)\right|^2 \quad (3)$$

Here *R* and *L* denote right and left electrodes. This expression is a sum over channels from orbital *m* to orbital *n*, but each channel contributes only positively to the conductance—that is, there is no interference between channels. The intensity of a given peak in the IETS spectrum is given by

$$W^\alpha = g_c \sum_{\substack{m\in\mathcal{L}\\ n\in\mathcal{R}}} \Gamma^L_{mm}(E)\Gamma^R_{nn}(E)\frac{1}{2}\left|\frac{\partial G_{mn}(E)}{\partial Q^\alpha}\right|^2 \quad (4)$$

***Application to Characteristic Molecular Topologies.*** A simple description is given in terms of a tight binding type, or extended Hückel type, model. This is a one-electron Hamiltonian, in which each interface orbital couples with the electrode, and the orbitals along the molecule couple only to nearest neighbors. We analyze three important specific cases—the linear chain, side chain structures, and planar conjugated organics.

***A. Linear Chain***. We consider first a linear chain of atoms with one orbital per atom. Calling the two terminal orbitals on the molecule 1 and *N*, the summations in Eqs. 3-4 can be limited to a single term containing $G_{1N}(E)$ and $\partial G_{1N}(E)/\partial Q^{\alpha}$ respectively. We write these Green's function matrix elements (Eq. 6) in a Dyson expansion:

$$G_{1N}(E) = \frac{(E - E_1)}{(E - E_1 - \Sigma_{11}^{L})(E - E_N - \Sigma_{NN}^{R})} \prod_{j=1,N-1} \frac{V_{j,j+1}(Q)}{(E - E_j)} \tag{5a}$$

$$\frac{\partial G_{1N}(E)}{\partial Q_\alpha} = \sum_{j=1,N-1} \frac{\partial V_{j,j+1}(Q)}{\partial Q_\alpha} \frac{G_{1N}(E)}{V_{j,j+1}(Q)} \tag{5b}$$

To interpret: $\Sigma^{L}$ and $\Sigma^{R}$, respectively, are self-energies in the right and left electrodes. This expression is just the familiar superexchange form,[16] modified to include self-energies. This highly intuitive expression says that for each channel between sites 1 and *N*, dependence upon the normal mode arises from variations in mixing matrix elements between neighboring orbitals, upon displacement along the given normal coordinate.

Unless the coordinate $Q_\alpha$ is totally symmetric, symmetrical linear chains exhibit very weak or vanishing IETS for that normal mode, because contributing terms of opposite signs will cancel in Eq. 5b. For the model system in Figure 1, the totally symmetric mode (a) contributes to IETS, while the non totally symmetric mode (b) will have effectively no IETS intensity.

***B. Side Chains***. For a molecule with a side chain the analysis becomes more complex. There might now be several significant paths for charge transport through the molecule. Eq. 7 then generalizes to

$$g(E) = g_c \sum_{\substack{m\in\mathcal{L} \\ n\in\mathcal{R}}} \Gamma_{mm}^{L}(E)\Gamma_{nn}^{R}(E) \left| \sum_{path\ P} \left[ G_{mn}^{P}(E) \right] \right|^2 \tag{6}$$

Here *P* denotes the different paths that the electron can travel between atomic orbitals *m* and *n*; the notation $G^P$ means the Green's function element along pathway *P* from *m* to *n* on the molecular bridge (expressed as in Eq. 5a). Intuitively, the right-hand side of the equation is quite obvious: the propagation through the molecule must go from site *m* to site *n*, but can do so through several different pathways among the electronic basis set—along each pathway, the derivative of the Green's function (Eq. 8) can occur on any of the mixing elements between two sites. The dominant pathways idea is reminiscent of the tunneling pathway scheme developed by Beratan.[18]

Clearly, the dominating pathways for transport through the molecule have the best combination of local overlap and small energy denominators. Electrons moving from one end to the other will not divert along paths that sample side chains very strongly—therefore, one would expect that normal coordinates dominated by side chain motion will contribute very weakly (e.g. Fig.1(c)-(d)).[19] Unlike propensity rule A, this rule is based not on symmetry, but rather on the very small probability for motion through the side chain. This is illustrated by the modest contribution of the CH stretching to the IETS spectrum.[6]

***C. Planar Conjugated Organics with $C_{2h}$ symmetry***. Many interesting molecular junctions involve conjugated planar molecules. In most cases (such as benzene dithiol, oligophenyleneethynylene, oligophenylenevinylene and their derivatives) the molecule has a $C_2$ axis perpendicular to the molecular plane (with overall point group $C_{2h}$) and symmetry considerations yield several useful propensity rules. We consider molecules attached to the electrode by a single atom (usually S attached to Au). In general all valence orbitals on the terminal atoms may be effectively coupled to the electrode, and the summation of Eq. 3 must be extended over the atomic orbitals *m* and *n* of the left and right sulfur atoms. Depending on the symmetry of the *m* and *n* orbitals and any eventual symmetry relation between them, group theory can predict which vibrational modes $\alpha$ lead to non- zero $\partial G_{mn}(E)/\partial Q^{\alpha}$ . We label *m* or *n* as σ-type or π-type if they are symmetrical or antisymmetrical upon reflection in the molecular plane. Moreover, if *m* and *n* are interchanged by the $C_2$ axis we say that the *m-n* couple is *symmetry related* (e.g. the 3s orbitals on the left and right sulfur are symmetry related). Results of group theoretical analysis can be summarized as follows:

(i) If *m* and *n* are both π type orbitals, the matrix element $\partial G_{mn}(E)/\partial Q^{\alpha}$ is non null for $a_g$ and $b_u$ modes. However, if *m* and *n* are symmetry related, only $a_g$ modes contribute to the corresponding matrix element. Since there is only one orbital of π type on the left and right sulfurs, and they are symmetry related, we conclude that only $a_g$ vibrations will contribute to the IETS through the π system. It then follows from B that one expects totally symmetrical C-C stretching and C-C-C bending to dominate in planar hydrocarbons.

(ii) If *m* and *n* are both σ type orbitals, $a_g$ and $b_u$ modes should give non null $\partial G_{mn}(E)/\partial Q^{\alpha}$ . But because σ-type transport through planar organics is much weaker than π-type transport, the overall IETS peaks from the σ channel is expected to be very weak.

(iii) If *m* and *n* are of different type —one σ the other π— there must be one spot along the pathway where motion along the normal coordinate $Q_\alpha$ permits π/σ mixing. These will be out-of-plane modes, with symmetry $a_u$ or $b_g$. Their IETS intensity will be strong if most of the tunneling path extends through the π system.

Note that, following these considerations, $b_u$ modes should *not* be observed in this class of compounds. Among the allowed modes, $a_u$ or $b_g$ will be localized in the lower frequency region of the spectrum (since they correspond to out-of-plane low-energy modes in conjugated organic molecule), while the allowed in-plane $a_g$ modes will dominate the high-frequency IET spectrum. These interesting propensity rules suggest that by comparing the IETS intensity with the vibrational spectrum of the isolated molecule, we may be able to deduce information from the IETS spectrum about the nature of the binding at electrode surfaces. Moreover, they correlate the activity of each vibrational mode to the type of tunneling (π-π, σ-σ or mixed) modulated by that mode.

***Some computational examples***. To evaluate the validity and utility of these propensity rules, we have calculated the expected IETS for a group of sulfur terminated molecules with $C_{2h}$ symmetry (shown in Fig. 2). We assumed that the S atoms are in an fcc-like adsorption site, with the S-Au distance of 2.85Å and the S-C bond perpendicular to the surface. A cluster containing 9 gold atoms was used only to compute the electrode-molecule interaction. We adopted the B3LYP density functional, the 6-31G* basis set for the organic part and the 'Lanl2mb' basis set/pseudopotential for Au.[20]

Fig. 1 presents the frequencies, intensities and symmetries calculated for the IETS spectrum of four important molecular bridges. It is demonstrated elsewhere[10] that DFT-calculated spectra based on the formal development of Eqns. 1-6 agree very well with experimental data, without any scaling or other adjustments (the actual line shape issue is more complicated[21,22]). Here we concentrate on the capacity of the propensity rules for describing the spectrum.

At high and medium frequencies (above ~800 $cm^{-1}$), the spectrum of the pi-electron species is dominated by vibrational modes of allowed $a_g$ symmetry (*in plane* C-C stretching and C-C-C bending). At lower energies we observe strong $a_u$ and $b_g$ modes in all four species – these arise largely from *out of plane* motions of the carbon backbone. Because no electrode atoms are included in the DFT normal mode calculation, these low-frequency features (allowed by the propensity rules) are purely molecular. According to the proposed propensity rules $b_u$ modes are forbidden if the tunneling takes place mainly through the π orbitals. In agreement with this rule $b_u$ modes are absent in the spectra of the three longer conjugated molecules and give a modest contribution only for the shorter molecule of the series (benzenedithiolate), for which the tunneling through the $\sigma$ orbitals is not completely negligible.

***Conclusions***. One can deduce useful propensity rules to describe the IETS spectrum. The assumptions necessary to arrive at these are straightforward, and hold in the Landauer-Imry tunneling regime, assuming the IETS spectrum is weak, that the bridge is bonded to the electrodes in a finite number of places, and that a simple tight binding model is adequate. Having used the normal modes of the isolated molecule (no metal atoms) all predicted spectral structure is due solely to motion of the molecule in the junction.

Selection rules, such as those seen in optical, infrared and Raman spectra, are not obtained for IETS[8,15] because IETS features arise not from the field-dipole interaction characterizing these other spectroscopies, but rather from the vibronic modification of the electronic levels. Expansion of the Landauer-Imry formula as a Taylor series in molecular normal coordinates gives a convenient, accurate perturbation-type formula for calculating both frequency and intensity of the IETS spectrum. Expansion in a Dyson-like form permits derivation of propensity rules, both symmetry-based and pathway-deduced. These permit a correlation of structure and coupling geometry with the IETS spectrum. These propensity rules work very well, for the calculated spectrum of four typical molecular bridges.

The broad propensity rule situations illustrated here can be extended to deal with specific structures. The wide interest in MJ transport is plagued by lack of structural direct information. These propensity rules suggest that the IETS spectrum provides useful geometric information about the actual molecular geometry in the junction.

Acknowledgments: We thank the DARPA MoleApps program, the NSF-MRSEC program and the Research Council UK for support.

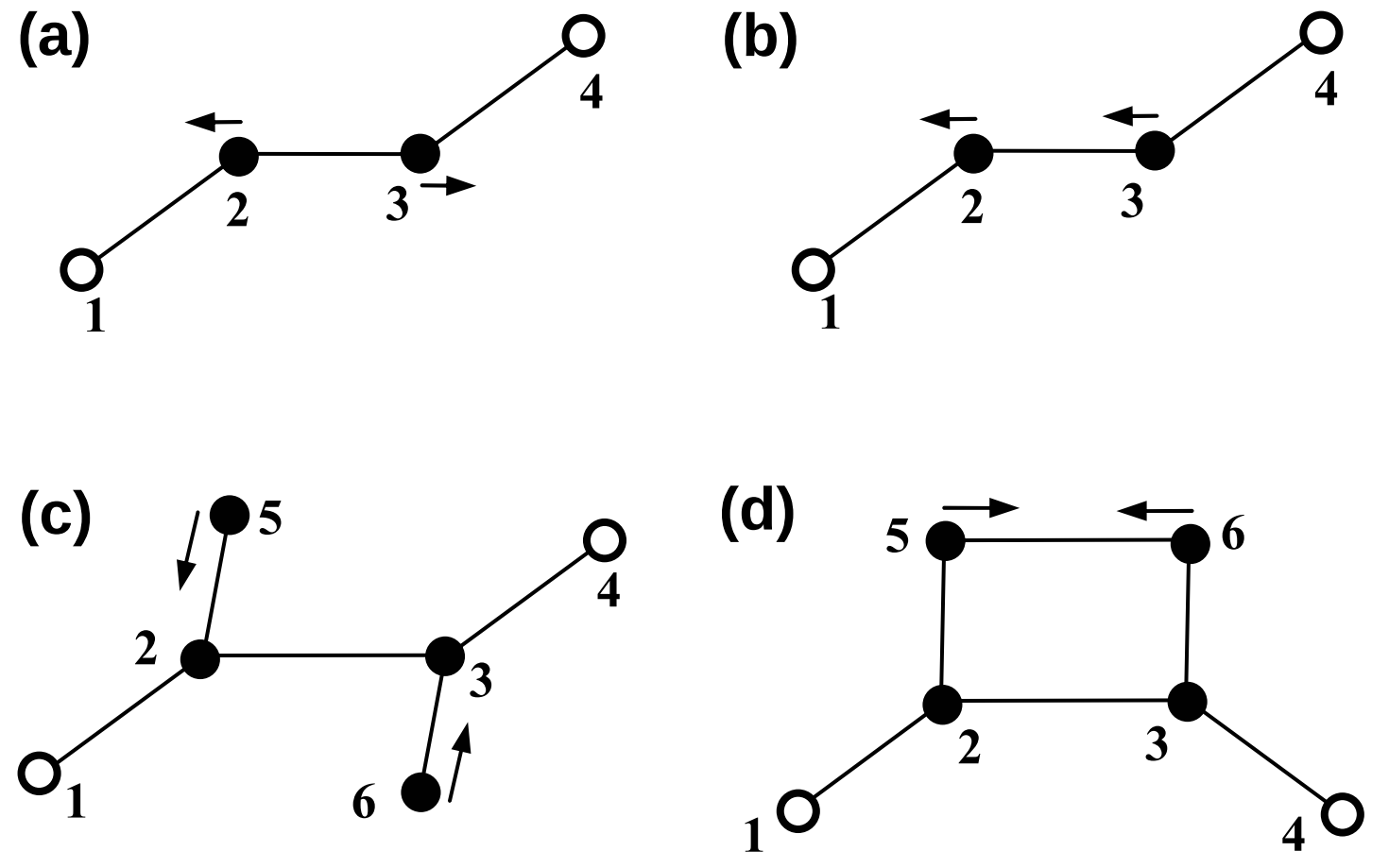

**Figure 1**. Idealized molecular wires represented as graphs. Each node (circle) represents an atomic orbital, empty circles are in contact with left and right electrodes, and interacting orbitals are connected by a line. The arrows indicate atomic motions in a normal mode. For the linear molecule (a)-(b) totally symmetric modes, like the one represented in (a), are IETS active, while non totally symmetric modes, like the one in (b), are forbidden. The totally symmetric modes represented in (c) and (d) are formally IETS allowed, but their actual contribution is small because they do not involve the main tunneling path 1-2-3-4.

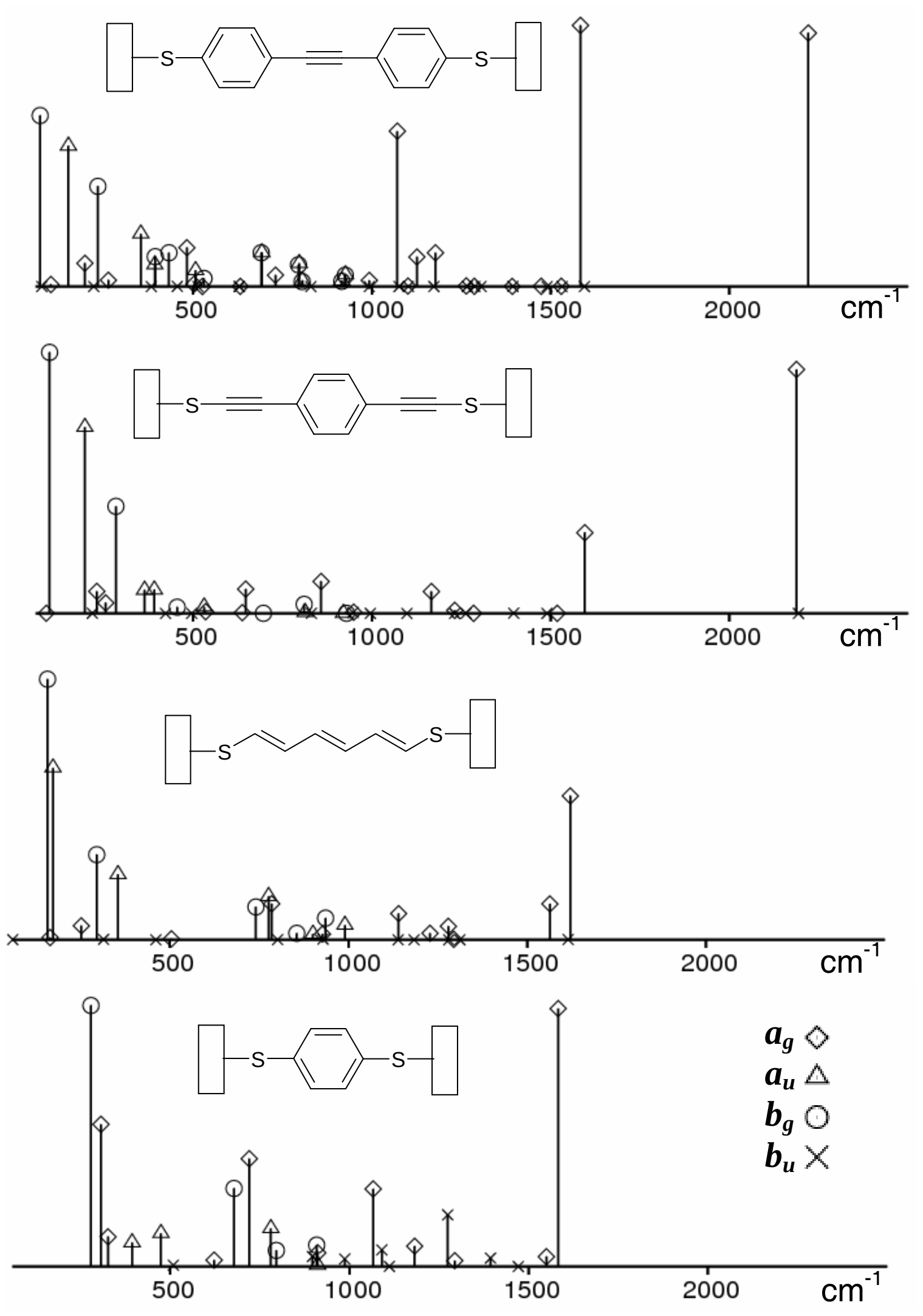

**Figure 2.** Simulated IETS spectra of several compounds of $C_{2h}$ symmetry. Bar heights are proportional to the peak intensity. The symbol on their top indicates the symmetry representation of the corresponding vibrational mode, as indicated in the legend. Note that only $a_g$ vibrations are active at high frequencies and mostly $a_u$ and $b_g$ are active at low energy; $b_u$ type vibrations are effectively forbidden for the three longer molecules of the series.